%% file: ML_Lat2022.tex
\documentclass[a4paper,11pt]{article}
\usepackage{pos}

\definecolor{darkcyan}{rgb}{0, 0.5, 0.5}

\title{Generative models for scalar field theories: how to deal with poor scaling?}

\author*[a]{Javad Komijani}
\author[a]{Marina K. Marinkovic}

\affiliation[a]{Institute for Theoretical Physics, ETH Zurich, 8093 Zurich, Switzerland}

\emailAdd{jkomijani@phys.ethz.ch}

\abstract{%
Generative models, such as the method of normalizing flows, have been suggested
as alternatives to the standard algorithms for generating lattice gauge field
configurations.
Studies with the method of normalizing flows demonstrate the proof of principle
for simple models in two dimensions.
However, further studies indicate that the training cost can be, in general,
very high for large lattices. The poor scaling traits of current models indicate
that moderate-size networks cannot efficiently handle the inherently multi-scale
aspects of the problem, especially around critical points.
We explore current models with limited acceptance rates for large lattices
and examine new architectures inspired by effective field theories to improve
scaling traits. We also discuss alternative ways of handling poor acceptance
rates for large lattices.%
}

\FullConference{%
The 39th International Symposium on Lattice Field Theory,\\
8th-13th August, 2022,\\
Rheinische Friedrich-Wilhelms-Universität Bonn, Bonn, Germany
}

\begin{document}
\maketitle

\section{Introduction}

The method of \emph{trivializing maps} was formulated by L\"uscher~\cite{Luscher:2009eq}
to improve the efficiency of Markov Chain Monte Carlo (MCMC) simulations of
lattice QCD by mapping the theory to another one that is easier to simulate,
ideally to a theory in which the degrees of freedom are decoupled.
L\"uscher discussed how to construct such a map systematically by integrating
certain flow equations in field space
and pointed out that, once such a map is constructed, the theory ``can be simulated
simply by generating uniformly distributed random gauge fields''~\cite{Luscher:2009eq}.
Although the last remark seemed ``likely to remain an academic one''~\cite{Luscher:2009eq},
it took less than one decade that a similar idea, which is called the method of
\emph{normalizing flows} (NF),
flourished with many applications such as image generation;
for review, see Refs.~\cite{Kobyzev:2021abc, Papamakarios:2021abc}.
The method of normalizing flows is implemented using deep neural networks rather
than integrating certain flow equations.
Deep neural networks can approximate a huge class of functions and,
as a result, provide a way to tackle complicated problems without a need to model
them first, in this case, constructing some flow equations and integrating them.
This, however, does not mean that one cannot use theoretical developments
to construct more suitable neural network architectures for NF.


Li and Wang~\cite{Li:2018nnrg} used a flow-based method for sampling from
a dual version of two-dimensional Ising model that resembles a scalar field
theory.
Albergo \emph{et.al.}~\cite{Albergo:2019eim, Albergo:2021vyo} extended the study
by applying NF to scalar field theories with quartic potential in two-dimensional
lattices up to $14^2$ sites
and discussed in detail different aspects of the algorithm
such as effects on the autocorrelation time.
Del Debbio \emph{et.al.}~\cite{DelDebbio:2021qwf} explored the scalability of the
method by investigating lattices up $20^2$ sites using different architectures.
Their study indicates that, in general, the method's efficiency deteriorates as
the lattice size increases (for a fixed architecture).
For a review of applications of generative models on lattice field theory,
we refer the reader to Ref.~\cite{Boyda:2022nmh}.
In this manuscript, we expand the study of scalar field theories with quartic
potential in two dimensions by introducing a novel flow model
inspired by effective field theories, discuss the scalability of the model,
and present a way to deal with the low acceptance rate at large volumes.

\section{Background and review of widely used architectures for NF}
\label{background}

\begin{figure}
    \vspace{-0.5cm}
    \includegraphics[width=\textwidth]{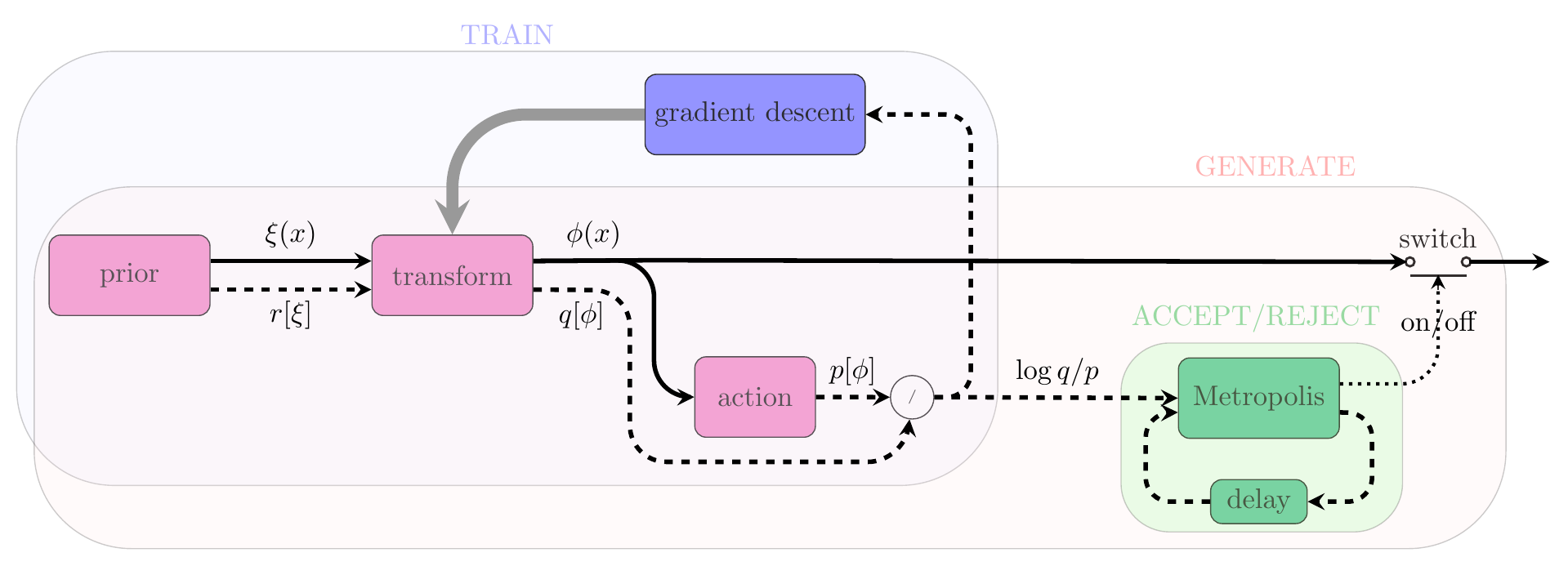}
    \caption{Block diagram for the method of normalizing flows.
    $\xi(x)$ and $\phi(x)$ are the prior and
    transformed fields at position $x$, and and $r[\xi]$ and $q[\phi]$ are
    corresponding probability densities. The ``GENERATE`` block illustrates
    the integration of NF and MCMC by including an accept/reject step.}
    \label{fig:NF:diagram}
\end{figure}

Let us start with a quick comment about the method of inverse transform sampling
(ITS).
This method can be used to draw samples from the probability density function
(PDF) of a continuous random variable, $f_Y(y)$,
by sampling from a simpler one, $f_X(x)$, and transforming the samples using
\begin{equation}
  y = F_Y^{-1} \circ F_X (x)\,,
\end{equation}
in which $F_X$ and $F_Y$ stand for the cumulative distribution functions of $x$
(the \emph{prior}) and $y$ (the \emph{target}) variables.
The method of NF can be considered a generalization of the ITS method
to higher dimensional distributions. With the method of NF, we deal with
a series of invertible and differentiable
transformations that are typically implemented by deep neural networks.
The series of transformations map the prior variable/distribution to a
new one that we simply refer to as the \emph{transformed} variable/distribution.
Training a NF-based model is then nothing but optimizing the parameters of the
model such that the transformed distribution resembles the target distribution.
To this end, one can minimize the relative entropy of the transformed and target
distributions using the Kullback-Leibler (KL) divergence
\begin{align}
   D_\text{KL}(q || p) &\equiv \int d\phi ~ q[\phi] \log \frac{q[\phi]}{p[\phi]} ~~ \ge ~0.
   \label{eq:KL}
\end{align}
Here, $\phi$ denotes the transformed variable;
$p[\phi]$ is the target PDF;
and $q[\phi]$, which can be written in terms of the prior PDF and the Jacobian
of transformation, is the transformed PDF.
The equality in KL divergence holds only if $p[\phi] = q[\phi]$.
The ``TRAIN'' block in Fig.~\ref{fig:NF:diagram} depicts the described
training procedure.
Here, $\xi(x)$ and $\phi(x)$ are the prior and transformed variables (fields) at position
$x$, and $r[\xi]$ and $q[\phi]$ are corresponding probability densities.
For the prior, we use a set of independent normal distributions.
The target PDF is 
$$
  p[\phi] = \frac{1}{Z} e^{-S[\phi]},
$$
where $S$ is the action of the theory and the normalization factor $Z$ is
typically not known, indicating that the lower bound in \eqref{eq:KL}
is not known.

Once the model is perfectly trained, one can use it to draw samples from the
target distribution.
In practice, however, it is unlikely to find a perfectly trained model,
especially when the degrees of freedom increase.
To correct the samples, one can integrate the method of NF with MCMC.
For example, Ref.~\cite{Albergo:2019eim} introduced an accept/reject
step as used in the Metropolis-Hastings algorithm
to ensure exactness.
The ``GENERATE'' block in Fig.~\ref{fig:NF:diagram} illustrates
such an integration, in which the accept/reject step is formulated using the
logarithm of the ratio of transformed and target densities,
$\log (q[\phi]/p[\phi])$, of consecutive proposed fields as input.

The method of normalizing flows requires invertible transformations, putting
some restrictions on NF architectures.
\emph{Coupling flows} are one of the most widely used architectures;
see Refs.~\cite{Kobyzev:2021abc, Papamakarios:2021abc} for reviews of different
types of flows.
With coupling flows, one divides the field degrees of freedom into two partitions,
which can be labeled as $a$ (active) and $f$ (frozen/fixed) partitions.
A checkerboard-like mask is convenient for such partitioning.
Each coupling-flow layer transforms the active partition by a function parametrized
with the frozen partition of the data:
$$
  x_a ~\to~ T(x_a ; \Theta(x_f)) \,.
$$
It is convenient to employ element-wise operations for $T$,
e.g., element-wise linear (affine) and spline transformations.
With such transformations, the Jacobian matrix is triangular,
making it easy to calculate its determinant.
Contrary to $T$, the form of $\Theta$ can be extremely complicated,
which is usually implemented by deep neural networks.

There are two widely used neural networks to model $\Theta$:
linear (dense) networks and convolutional networks.
The former is great for small-size lattices, but the number of parameters
grows fast as the size of the lattice grows.
The latter takes advantage of the translational symmetry of the underlying
theory and in general needs much fewer parameters. However, the latter
requires many layers of neural networks to propagate the correlation
throughout the data.

\section{Designing new architectures for normalizing flows}

\subsection{Effective action and power spectral density}

Inspired by symmetries of the action and effective theories of scalar fields,
our primary goal in this section is to construct a novel flow layer that
can propagate the correlation in data in an efficient way.
To this end, we start with an effective description of a real, scalar field
theory. The action of such a field in $d$ spacetime dimensions is
\begin{align}
  S[\phi] = \int d^d x
    \left(\frac{\zeta}{2} \partial_\mu \phi(x) \partial_\mu  \phi(x)
     + \frac{m^2}{2}  \phi^2(x)
     + \sum_{n=3}^\infty g_n \phi^n(x) \right)\, .
  \label{eq:def:scalar-action}
\end{align}
The corresponding quantum effective action reads
\begin{align}
  \Gamma[\phi] = \frac{1}{2} \int \frac{d^d k}{(2\pi)^d}\,
           \Big(\zeta k^2 + m^2 + \Pi(k^2) \Big) |\tilde \phi(k)|^2
           + \cdots\,,
  \label{eq:effective:action}
\end{align}
where $\tilde \phi(k)$ is the scalar field in Fourier space.
The quantum effective action has the following property:
the tree-level Feynman diagrams that it generates give the complete scattering
amplitude of the original theory~\cite{Srednicki:2007qft}.
Note that $\Big(\zeta k^2 + m^2 + \Pi(k^2) \Big)$ is the inverse of the two-point
correlator and, employing the engineering terminology, it is proportional
to the inverse of power spectral density (PSD) generalized to $d$ dimensions.

\begin{figure}
   \begin{center}
       \includegraphics[width=7cm]{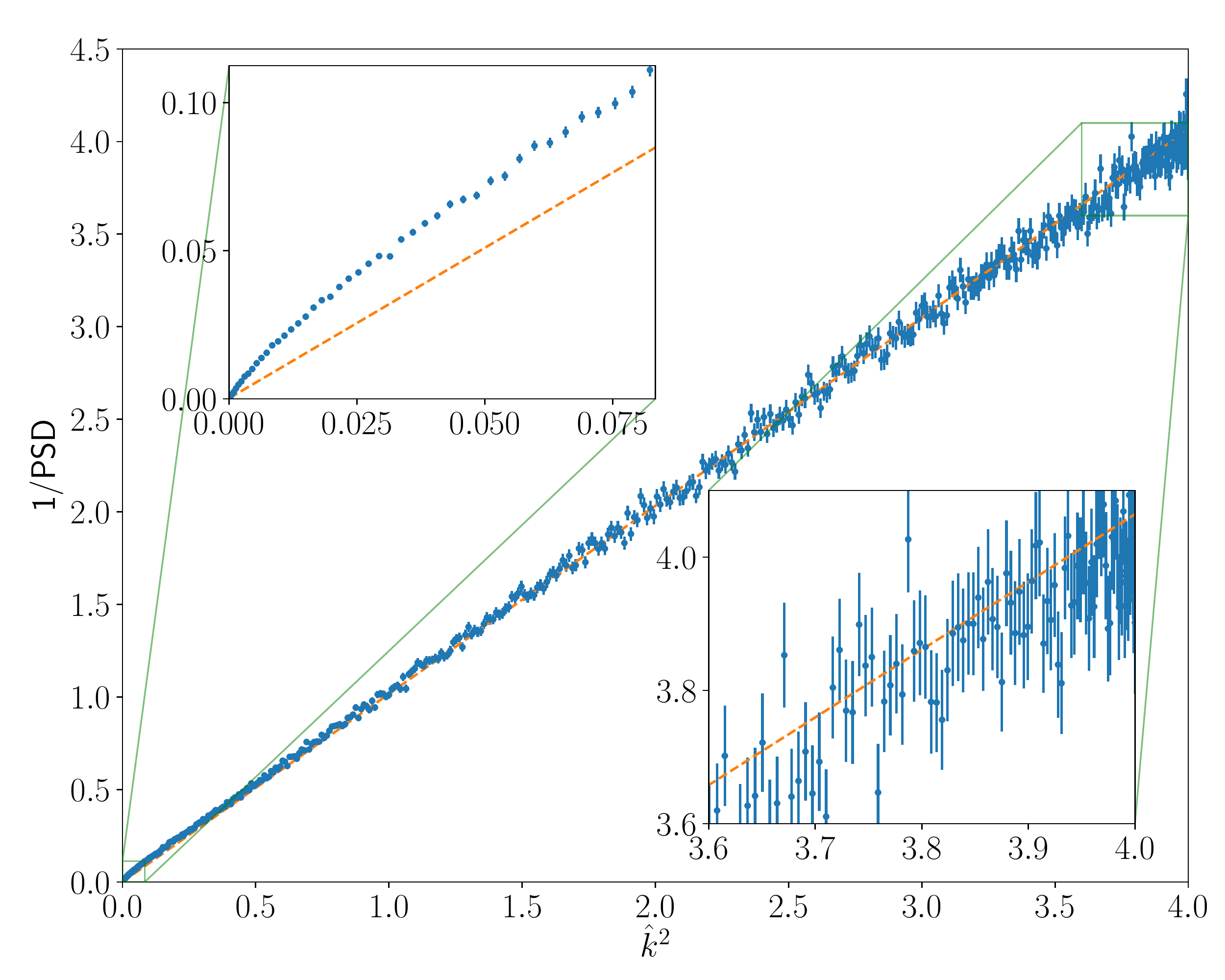}
       \includegraphics[width=7cm]{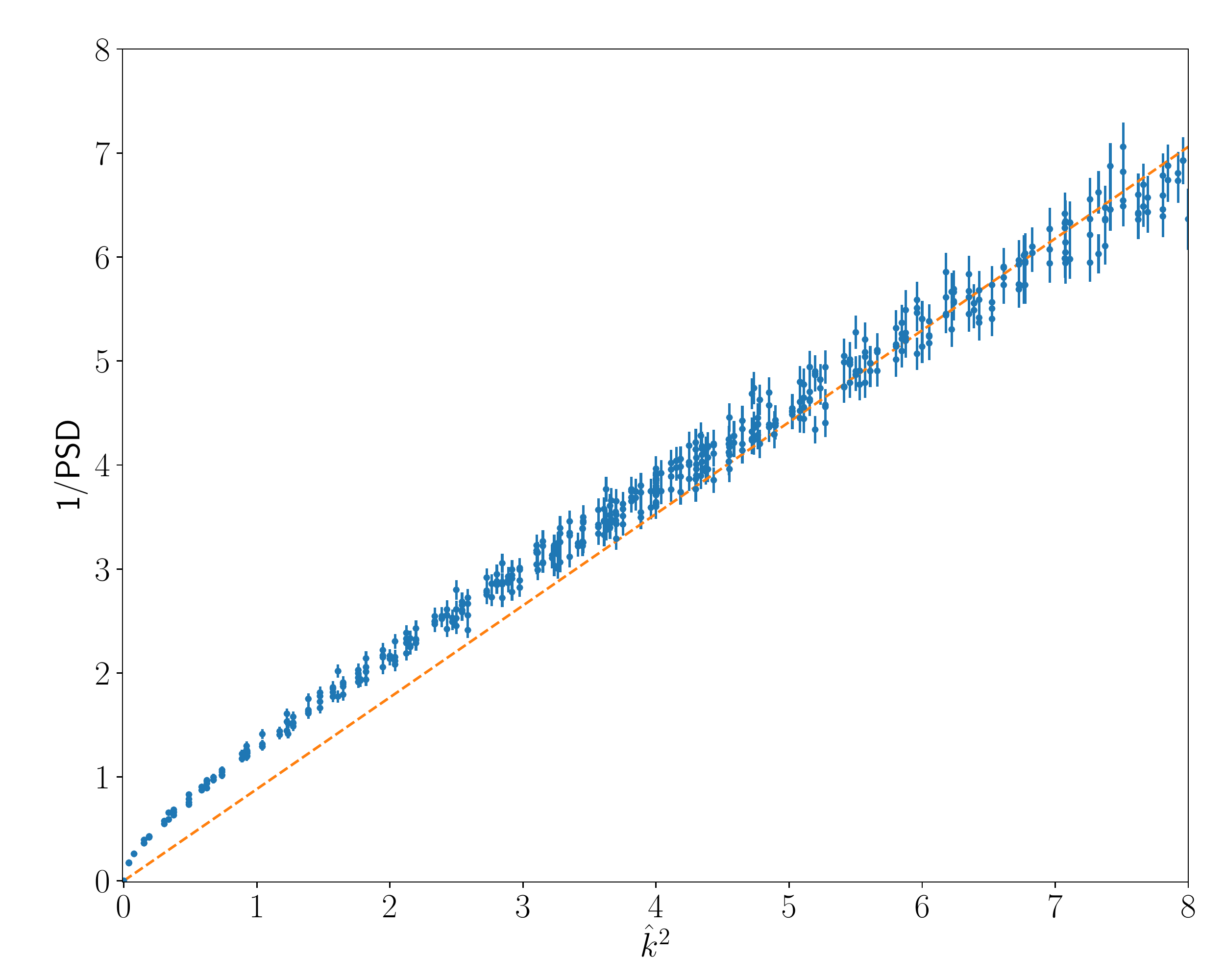}
   \end{center}
   \vspace{-0.5cm}
   \caption{The inverse of PSD of quartic scaler field theories
   with double-well potential (from MCMC simulations).
   Left: One-dimensional lattice with size $L = 1024$ and parameters $\zeta = 1$,
   $m^2 = -1.6$, $g_4 = 0.1$ (with lattice spacing set to $0.125$).
   Right: Two-dimensional lattice with size $L^2 = 32^2$ and parameters $\zeta=0.7$,
   $m^2 = -2.8$, and $g_4 = 0.5$.
   The dashed lines are guide for the eye.
   }
   \label{fig:IPSD:examples}
\end{figure}

As manifested in \eqref{eq:effective:action},
an element-wise operation on $\tilde \phi(k)$ can map the PSD to the one
of interest. Depending on the properties of PSD, one can restrict the map even
further. For example, the Lorentz invariance of PSD implies that
the element-wise operation depends only on $k^2$.
We now examine a couple of examples for further restrictions.
Figure~\ref{fig:IPSD:examples} shows the inverse of PSD of a $\phi^4$ scalar
theory with double-well potential in one and two
dimensions obtained from MCMC simulations plotted
against $\hat k^2 = \sum_i 4 \sin^2 (k_i/2)$.
The figure indicates that the inverse of PSD can be modeled using a positive,
monotonically increasing function of $\hat{k}^2$.
Here, we model the inverse of PSD with a rational quadratic spline (RQS)
\cite{Gregory:1982rqs, Delbourgo:1983rqs, Durkan:2019rqsf}
as a function of $\hat k^2$, and we scale $\tilde \phi(k)$ accordingly.

In the case of a two-dimensional problem, PSD blows up at $k^2 = 0$ in the
broken phase.
This special point can be handled using the mean field theory:
the mean-field potential turns to a double-well potential at the broken phase.
Therefore, at $k^2 = 0$, instead of scaling $\tilde \phi(0)$,
we feed it to a separate RQS, which can change the distribution
of $\tilde \phi(0)$ to a multi-modal distribution.

We use the term \emph{PSD flow} to denote the described transformation.
Note that a PSD flow can change the correlation in data at the largest
and shortest scales. In the next part, we investigate an architecture with one
PSD-flow layer followed by four coupling-flow layers.

\subsection{A new architecture}

In this part, we explain how we use a PSD-flow layer to construct a new
NF architecture for a real, scalar field theory in two dimensions,
and we investigate the scalability of the new architecture.

The architecture that we investigated contains three blocks.
A PSD-flow layer, followed by two blocks of affine coupling flows,
each block has two layers alternating the active and frozen partitions.
(In total, there are four coupling-flow layers.)
For the $\Theta$ function in the affine coupling flows, we use convolutional
neural networks.
Each of the three blocks has its own activation:
symmetric RQS, tanh, symmetric RQS, respectively.
Unlike the tanh activation, the symmetric RQS splines that we use have free
parameters.
We use symmetric RQS activations because they respect the $Z_2$ symmetry of
$\phi^4$ scalar theories.
In total, there are about 3.4~K parameters in the model.
We use this NF model for $\phi^4$ scalar fields in two dimensions
with several values of $L$: $8, 12, 16, \cdots, 64$.
We train the model for the $8^2$ lattice with 10~K epochs.
For the $12^2$ lattice, instead of training from scratch, we rely on transfer
learning:
we start from the model trained for the $8^2$ lattice
and train it for 5~K epochs.
Then, we use the model trained for the $12^2$ lattice as the starting point
for the $16^2$ lattice and analogously for larger lattices.

\begin{figure}
   \begin{center}
   \includegraphics[trim=0.6cm 0 0.5cm 0, clip, width=5.4cm]{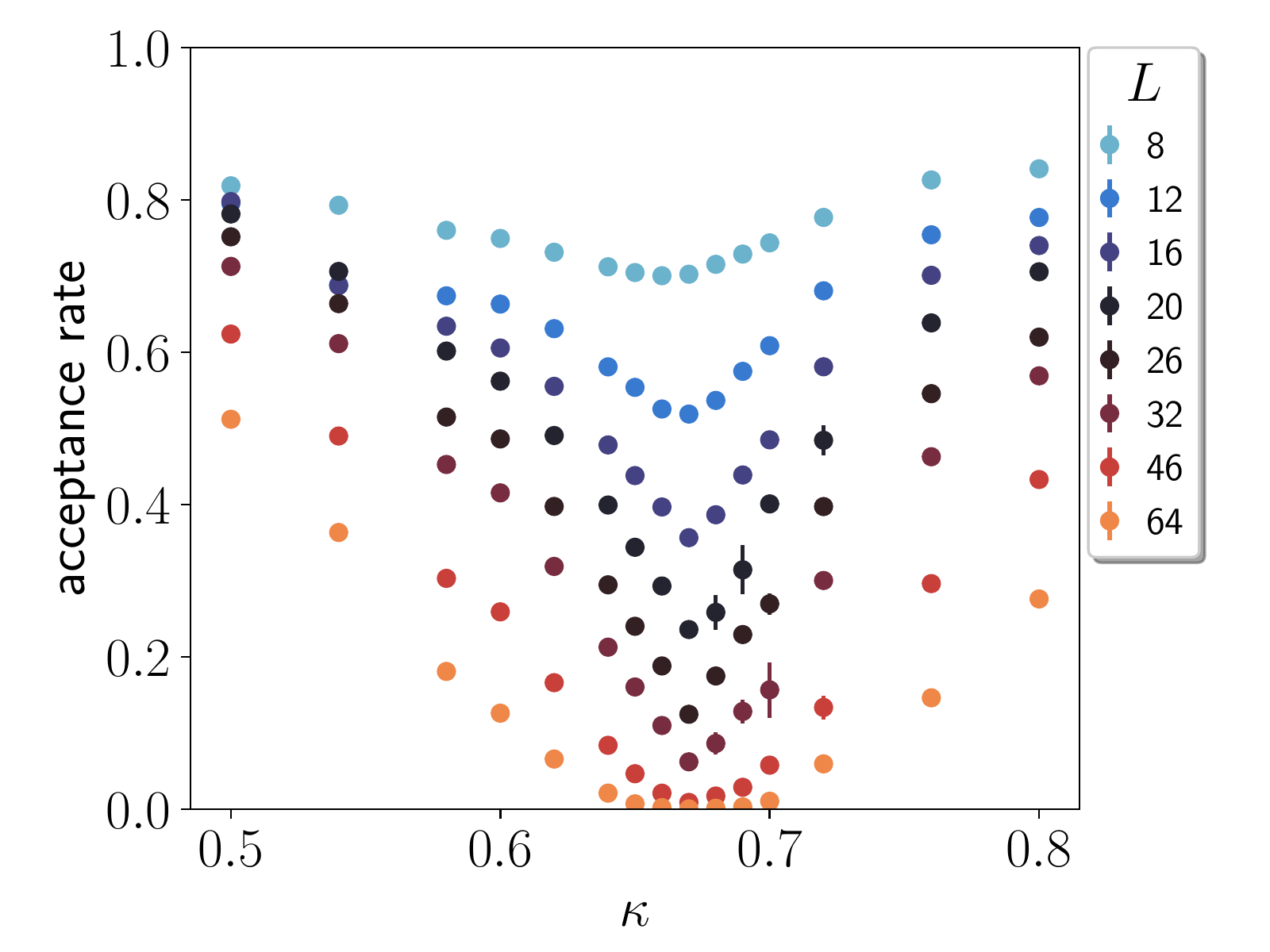}
   \includegraphics[trim=0.3cm 0 2.2cm 0, clip, width=4.7cm]{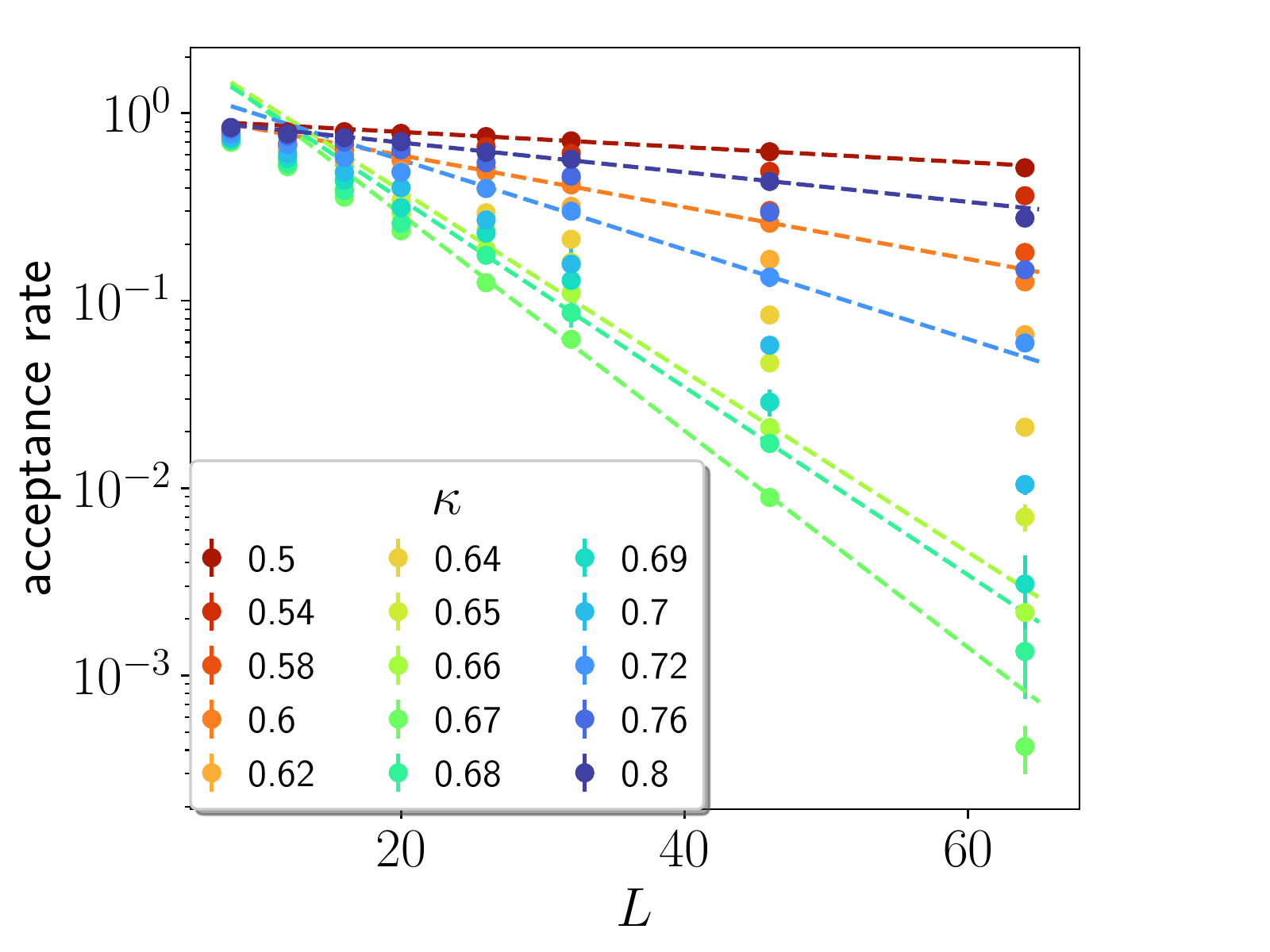}
   \includegraphics[trim=0.3cm 0 2.2cm 0, clip, width=4.7cm]{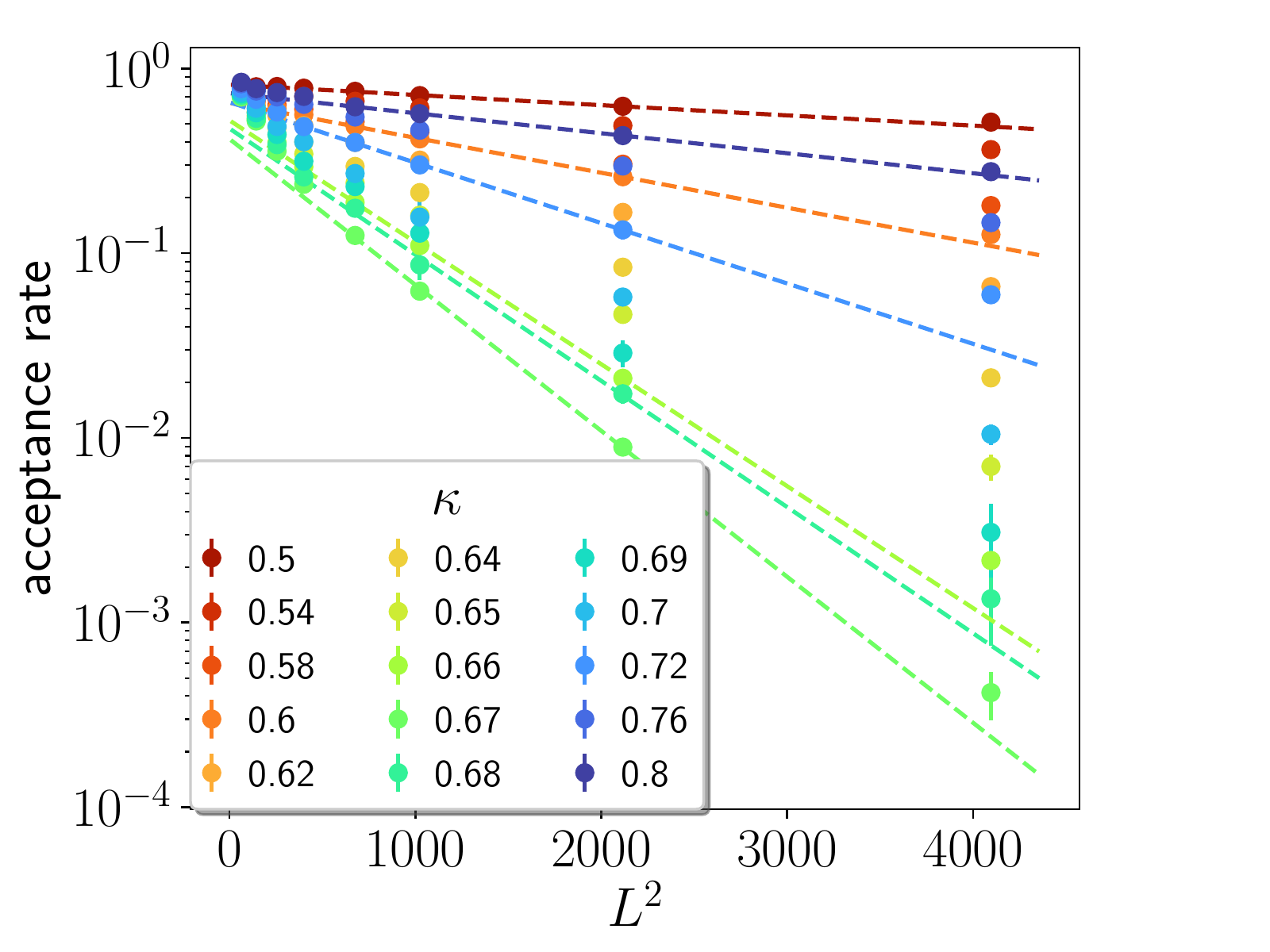}
   \vspace{-0.2cm}
   \caption{Acceptance rate plotted against $\kappa$, $L$, and $L^2$.
   The dashed lines are guide for the eye.
   }
   \label{fig:accept-rate}
   \end{center}
\end{figure}

In order to compare our results with the literature,
we fix the parameters of the action in \eqref{eq:def:scalar-action}
as follows: $\zeta = \kappa$, $m^2 = -4 \kappa$, $g_4 = 1/2$,
and $g_n = 0$ for $n \neq 4$.
Varying $\kappa$ from 0.5 to 0.8, we can compare our results with
Ref.~\cite{DelDebbio:2021qwf}[Fig.~4].
The left panel of Fig.~\ref{fig:accept-rate} shows the acceptance rate plotted
against $\kappa$ for several values of lattice size.
For $L=8$ lattices, the acceptance rate of the trained models has a mild dependence
on $\kappa$.
As the lattice size increases, the acceptance rate decreases.
Similar to Ref.~\cite{DelDebbio:2021qwf}, we observe that as $\kappa$ approaches
its critical value ($\kappa_c \approx 0.67$), the acceptance rate deteriorates
faster as the lattice size increases.

The middle and right panels of Fig.~\ref{fig:accept-rate} show the acceptance
rate plotted against $L$ and $L^2$, respectively.
The acceptance rate drops exponentially fast as $L$ increases,
but the asymptotic dependence cannot be reliably extracted from the graphs.
To investigate this behavior, we examine the acceptance rate and its dependence
on the distribution of $\log q[\phi]/p[\phi]$ by introducing a toy model in
the next part.

\begin{figure}
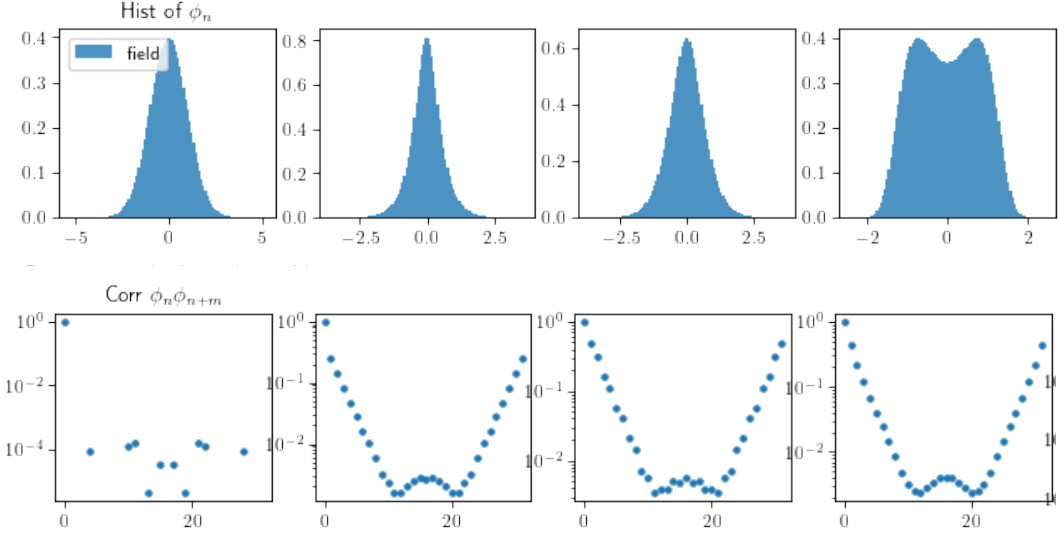

   \vspace{-0.2cm}
   \begin{center}
   \includegraphics[trim=0 10cm 5.1cm 0, clip, width=14cm]{figures/all-hist_b0.60_L32.pdf}
   \includegraphics[trim=0 0 5.1cm 10cm, clip, width=14cm]{figures/all-hist_b0.60_L32.pdf}
   \vspace{-0.2cm}
   \caption{Histograms of snapshots of $\phi(x)$ from the prior and outputs
   of three blocks of transformations.
   The lower panels show corresponding 2-point correlation functions.
   }
   \label{fig:example:hist:32}
   \end{center}
\end{figure}

Before concluding this part,
it is in order to examine the effect of employing a PSD-flow layer.
To this end, we present results from the model with parameters $\kappa = 0.6$
and $L = 32$.
Figure~\ref{fig:example:hist:32} shows histograms of snapshots of $\phi(x)$
from the prior (upper left panel)
and the outputs of three blocks of transformations (2nd, 3rd, and 4th columns,
respectively).
The lower panels show corresponding 2-point correlation functions.
We observe that the PSD-flow block (the second column) can introduce a correlation
to the data that roughly remains unchanged in the next blocks.

\section{Variance in $\log(q/p)$, acceptance rate, and poor scaling at large volumes}

The distribution of $\log (q[\phi]/p[\phi])$ determines the acceptance rate of
the model. For the architecture investigated here, we observe that
the variance of $\log(q/p)$ roughly scales with the volume of the lattice
in most cases.
This rough behavior can be heuristically explained as follows.
One can divide a large lattice into multiple blocks.
If the blocks are large enough, the field fluctuations in one block can be
considered independent of other blocks.
Then, the variance of $\log(q/p)$ is proportional to the number of blocks
and, in turn, to the volume of the lattice.

It is easy to compose models that yield a large acceptance rate for a small
lattice. As the lattice volume increases, the given model reaches a poor scaling
area. One can improve the model's performance by changing the
hyperparameters, adding more layers, or using more complicated architectures.
As an alternative approach, we introduce and use a method that we
call \emph{block updating}.
To this end, we first introduce a toy model, investigate it,
and explain the block-updating approach.

\subsection{Toy model}
\label{sec:toy}

\begin{figure}
   \begin{center}
   \includegraphics[trim=0.3cm 0 10.4cm 0, clip, width=4.6cm]{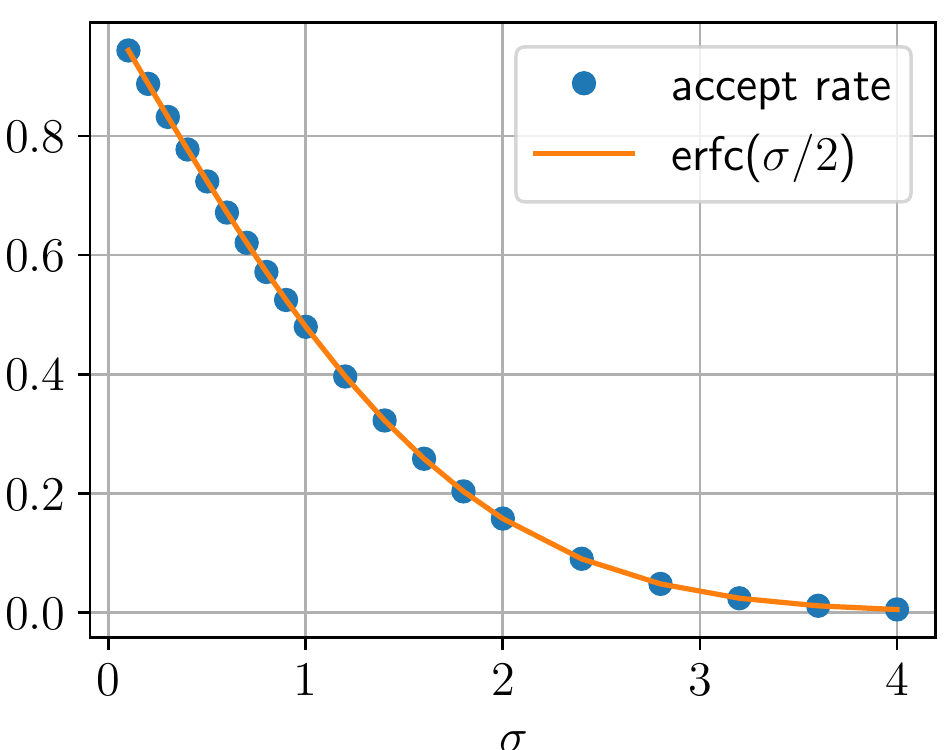}
   \includegraphics[trim=0.5cm 0 11.4cm 0, clip, width=5.1cm]{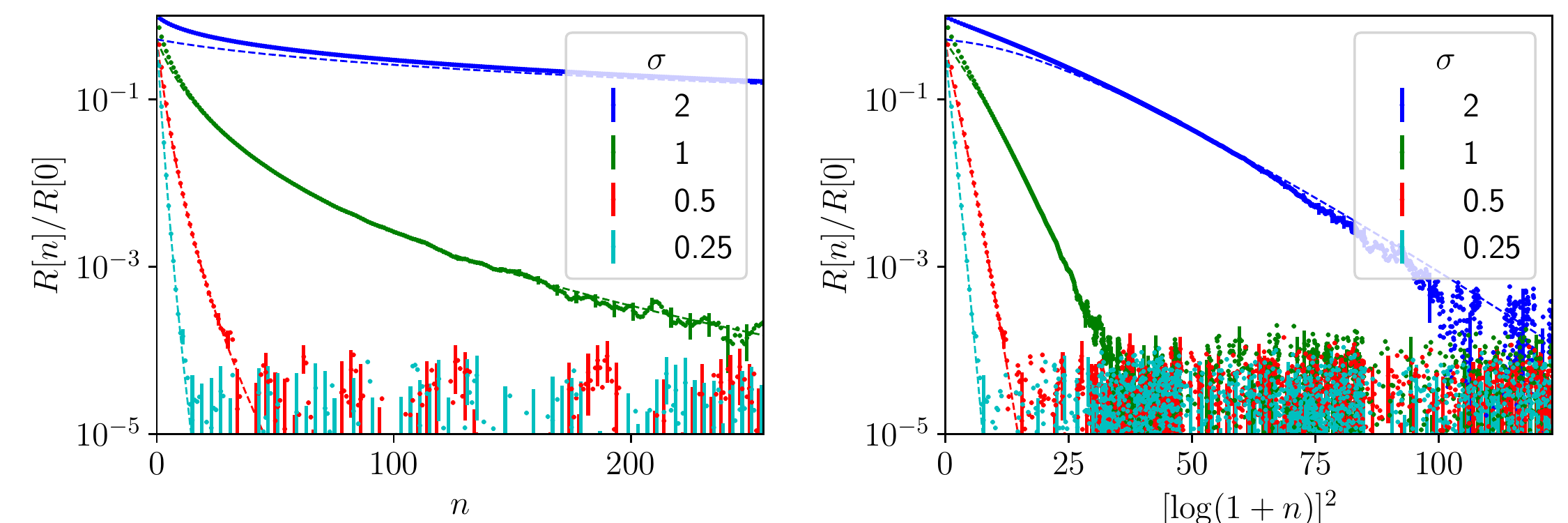}
   \includegraphics[trim=0.5cm 0 11.4cm 0, clip, width=5.1cm]{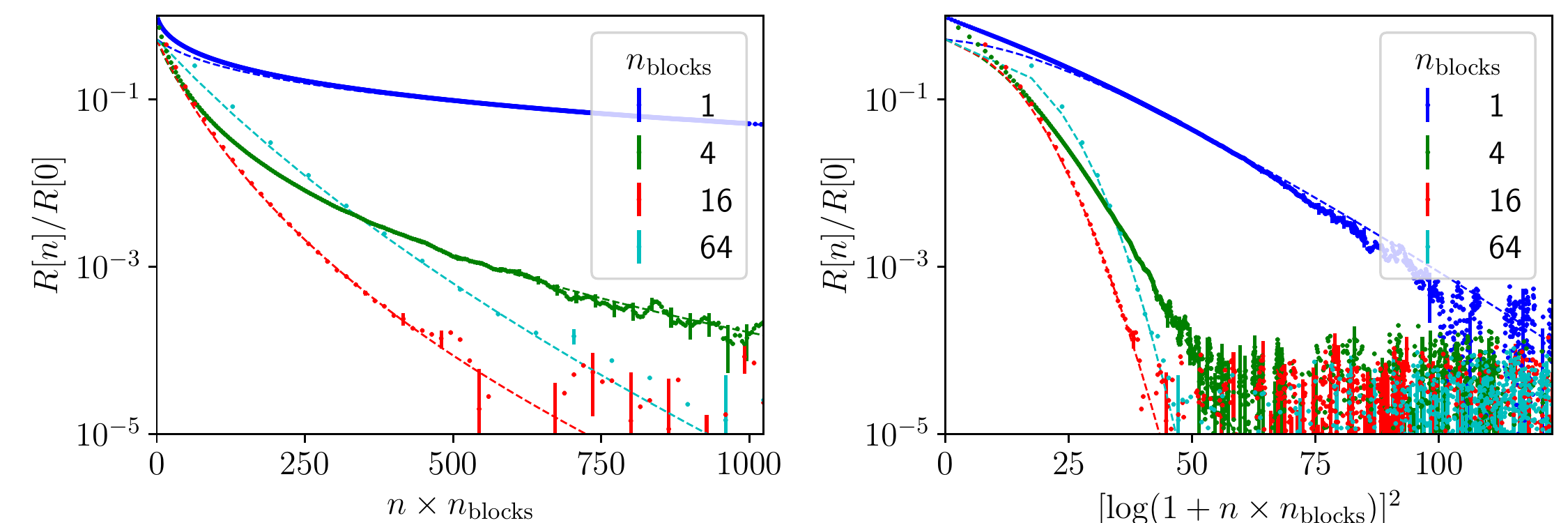}
   \caption{Acceptance rate (left panel) and autocorrelation function
   (middle and right panels) for the toy model introduced in Sec.~\ref{sec:toy}.
   The dashed lines show the asymptotic expression of the autocorrelation
   function.
   In the right panel, the block-updating procedure is used for several numbers
   of blocks and $\sigma=2$.
   }
   \label{fig:toy-model}
   \end{center}
\end{figure}

Let $x_{n}$ be a sequence of independent and identically distributed (iid)
random variables with normal distribution $\mathcal{N}(0, \sigma^2)$.
We define a new random sequence based on the Metropolis-Hastings accept/reject
step as
\begin{equation}
  \label{eq:def:JK-process}
  y_n = h(x_n, y_{n-1}) =
  \begin{cases}
   x_n\quad \text{with probability}~ e^{-\text{Relu}(x_n - y_{n-1})}\, \\
   y_{n-1}\quad \text{otherwise}
  \end{cases}
\end{equation}
for $n > 0$ and $y_0 = x_0$.
The conditional probability distribution, for $n > 0$, is
\begin{equation}
  \label{eq:cond-pdf}
  f_{Y_n | X_n , Y_{n-1}}(y_n | x_n , y_{n-1}) = \delta(y_n - x_n) e^{-\text{Relu}(x_n - y_{n-1})} 
          + \delta(y_n - y_{n-1}) \left(1 - e^{-\text{Relu}(x_n - y_{n-1})}\right).
\end{equation}
We are interested to calculate the (static) distribution of the $y_n$ sequence
for large values of $n$.
From
\begin{align}
   f_{Y_n}(y_n)
   &= \int dx_n\, dy_{n-1} f_{X_n}(x_n) f_{Y_{n-1}}(y_{n-1}) f_{Y_n | X_n , Y_{n-1}}(y_n | x_n , y_{n-1})\,,
\end{align}
we conclude that $Y_n \sim \mathcal{N}(-\sigma^2, \sigma^2)$ for large $n$.
The acceptance rate is then
\begin{align}
  \int dx_n dy_{n-1} f_{X_n}(x_n) f_{Y_{n-1}}(y_{n-1}) e^{-\text{Relu}(x_n - y_{n-1})}
  = \text{erfc}(\sigma/2)\,.
  \label{eq:toy:accept-rate}
\end{align}
The left panel in Fig.~\ref{fig:toy-model} illustrates $\text{erfc}(\sigma/2)$
and also the simulation values of acceptance rate plotted against $\sigma$.
For later use, let us calculate the asymptotic form of the acceptance rate.
From the asymptotic behavior of the complementary error function,
%
we conclude that for large $\sigma$,
\begin{equation}
    -\log(\text{accept rate}) = \frac{1}{4}\sigma^2 + \text{O}(\log(\sigma))\,.
    \label{eq:accept-rate:asymptotic}
\end{equation}

We now study the autocorrelation in the $y_n$ sequence defined as
$R[n]/R[0]$, with
\begin{align}
  R[n] &=  
      \mathbb{E} \left(y_k + \sigma^2\right) \left(y_{n+k} + \sigma^2\right)
\end{align}
for $k$ large enough.
The autocorrelation function can be calculated asymptotically for large $n$;
the expression is lengthy, and we do not reproduce it here.
The middle panel in Fig.~\ref{fig:toy-model} shows the
autocorrelation in $y_n$ for several values of $\sigma$.
The decay of the autocorrelation function is sub-exponential,
in agreement with the corresponding asymptotic expression shown by dashed lines.
%
%
For a fraction of points, rough estimates of uncertainties in determining
the autocorrelation are shown with error bars.

We aim to modify the model to decrease the autocorrelation in the $y_n$ sequence.
We implement a method that we call block updating.
First, we assume that $x_n$ is obtained by adding $n_\text{blocks}$ iid
normal variables with mean $0$ and variance $\sigma^2/n_\text{blocks}$ as
%
   $x_n = \sum_{b=1}^{n_\text{blocks}} x^{\{b\}}_n $.
%
Similarly, we decompose $y_n$ as
%
   $y_n = \sum_{b=1}^{n_\text{blocks}} y^{\{b\}}_n $.
%
Then, instead of proposing independent values of $x_n$ at each step,
we divide each step to $n_\text{blocks}$ substeps.
At each substep, we draw a new value for one block of $x_n$, i.e.,
for $x^{\{b\}}_n$, and propose it to update $y^{\{b\}}_n$:
\begin{equation}
  \label{eq:def:JK-process:blocked}
  y^{\{b\}}_n = h \left(x^{\{b\}}_n, y^{\{b\}}_{n-1}\right).
\end{equation}
Because the blocks are independent, the problem can be reduced to having
$n_\text{blocks}$ independent copies of the original problem with reduced
variance $\sigma^2/n_\text{blocks}$.
As the number of blocks increases, the reduced variance decreases and
consequently the acceptance rate increases (at the price of splitting each step
into $n_\text{blocks}$ substeps or having $n_\text{blocks}$ copies).
From equation \eqref{eq:accept-rate:asymptotic} one may conclude that
we do not gain any advantages because the probability of getting a completely
fresh vector $(y^{\{1\}}_n, \cdots, y^{\{n_\text{blocks}\}}_n)$ compared to
$(y^{\{1\}}_{n-1}, \cdots, y^{\{n_\text{blocks}\}}_{n-1})$,
in which all blocks are replaced with proposed ones,
does not change asymptotically because
\begin{equation}
    n_\text{blocks} \log 
           \text{erfc} \left(\frac{\sigma}{2\sqrt{n_\text{blocks}}}\right)
    = \sigma^2 + \text{O}(\log(\sigma))
\end{equation}
when $\sigma^2/n_\text{blocks}$ is large enough.
However, the block-updating procedure has a significant effect on the
autocorrelation in $y_n$.

There are two competing aspects in the block-updating procedure.
On the one hand, as $n_\text{blocks}$ increases, the outputs of consecutive
substeps get more correlated because we update only a block of the data at each
substep.
On the other hand, the acceptance rate increases for each substep,
which in general reduces the autocorrelation in the output.
The effects of these two competing aspects can be seen in the right panel of
Fig.~\ref{fig:toy-model}, which illustrates the autocorrelation in
$y_n$ with $\sigma = 2$ for several blocks: 1, 4, 16, 64.
In this panel,
to take into account the cost of block updating, i.e., splitting each step into
$n_\text{blocks}$ substeps, the argument of the autocorrelation function
(the horizontal axis) is inflated by the number of blocks.
We observe that
the decay of the autocorrelation function speeds up as the number of blocks
increases from 1, indicating that the effects of the second aspect are dominant.
But, after a certain point, the effects of the first aspect dominate and
autocorrelation time increases.
We leave detailed discussions on the integrated autocorrelation time
to future work.

\subsection{Variance of $\log(q/p)$, block size, and acceptance rate}

There are similarities and differences between the toy model introduced in the
previous part and the main problem investigated in this manuscript.
Assuming the distribution of $\log(q/p)$ is normal, one could identify
the sequence of proposed values of $\log(q/p)$ with $x_n$ in the toy model and
the sequence of accepted values of $\log(q/p)$ with $y_n$.
Then, one could apply the results of the previous section to study $\log(q/p)$
and, to some extent, other quantities.
There are three main differences.
Firstly, the distribution of $\log(q/p)$ is not necessarily normal.
Secondly, all quantities do not necessarily suffer from the same autocorrelation
in the sequence of accepted values of $\log(q/p)$.
Finally, the effects of applying a block updating procedure cannot be reduced
to having $n_\text{blocks}$ independent copies of a similar problem.

We first examine the relation between acceptance rate and volume.
As mentioned above, for the architecture studied here, we observe that
the variance of $\log(q/p)$ roughly scales with the volume of the lattice
in most cases.
Based on this observation and assuming the distribution of $\log(q/p)$ is normal,
one can employ the asymptotic relation in~\eqref{eq:accept-rate:asymptotic}
and argue that as $V\to\infty$,
\begin{equation}
    -\log(\text{acceptance rate}) \propto V + \text{O}(\log(V))\,.
    \label{eq:accept-rate_V}
\end{equation}
In practice, however, the above assumptions are not completely correct,
and by comparing the middle and right panels of Fig.~\ref{fig:accept-rate},
one may conclude that dependence of the logarithm of the acceptance rate
on the volume is milder than what equation \eqref{eq:accept-rate_V} suggests
for large volumes.
Even in some cases, the dependence looks more consistent with scaling by $\sqrt{V}$
rather than $V$, but this might be because the volume is not large enough
to use the asymptotic relation.
Moreover, note that these observations may change once one varies the settings,
e.g., by using a different model or increasing the number of epochs.

\begin{figure}
   \begin{center}
   \includegraphics[width=7cm]{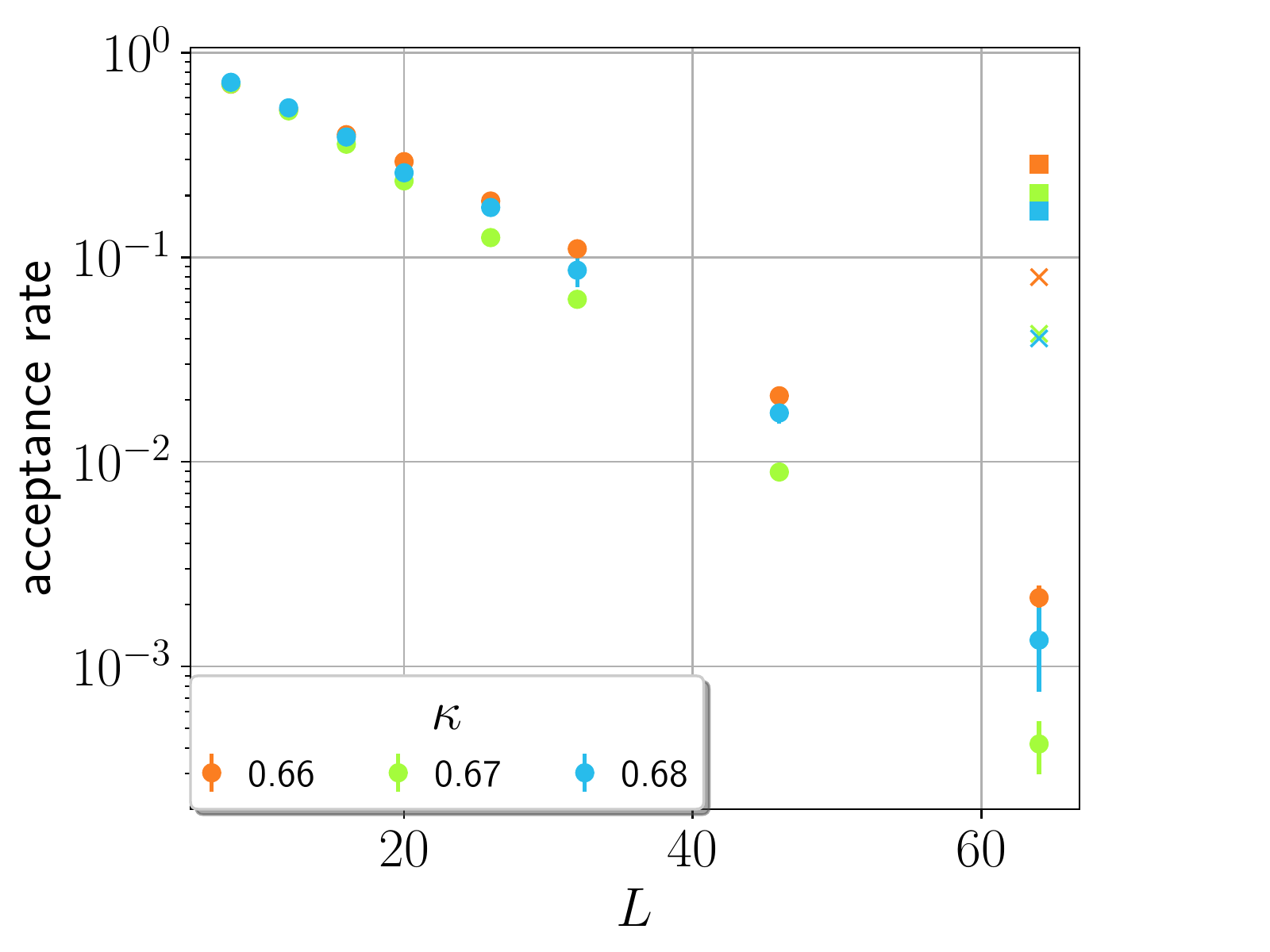}
   \caption{
   Effects of block updating on acceptance rate for three values of $\kappa$
   close to the critical point.
   The circle, cross, and square points show the acceptance rate for 1, 4,
   and 16 blocks applied on the largest lattice.}
   \label{fig:accept-rate-block-updating}
   \end{center}
\end{figure}

Similar to the toy model, we can use the block-updating procedure to improve
acceptance rate and integrated autocorrelation time.
To this end, instead of proposing completely independent configurations
at each step, we split the lattice into several blocks,
and at each substep, we update only the prior fields on the corresponding
block.
Figure~\ref{fig:accept-rate-block-updating} shows the effect
of block-updating procedure applied on the largest lattice, $L^2 = 64^2$,
for three values of $\kappa$ close to the critical point of theory.
The circle, cross, and square points show the acceptance rate
for 1, 4, and 16 blocks, respectively.
As expected, the acceptance rate improves as we increase the number of blocks.

Our primary investigation shows that the block-updating procedure introduced
here also improves the autocorrelation in various quantities.
We leave this discussion to another work.

\section{Summary and outlook}

In this manuscript, we reviewed coupling flows as one of the widely
use building blocks to construct NF architectures.
Inspired by effective field theories,
we presented a new transformation called PSD flow.
With a new architecture that employs a PSD-flow layer
and (in total) 4 coupling-flow layers, we investigated lattices up to $64^2$ sites.
Although the new architecture allows us to increase the lattice size,
the model's acceptance rate deteriorates at large volumes
in a fashion similar to what was observed in Ref.~\cite{DelDebbio:2021qwf}.

To investigate the behavior of the acceptance rate as a function of the volume
of the lattice, we introduced a toy model and discussed how one could handle the
poor acceptance rate and long integrated autocorrelation time of the toy model
by block updating.
Based on the similarities between the toy model and the problem at hand,
we proposed that a block-updating procedure can be employed to handle the poor
scaling of the acceptance rates for large lattices.

We are extending our studies to other theories, e.g., gauge theories, and
applying the PSD flow to these theories.
We are also exploring variants of the PSD flow.
Moreover, we are investigating the effects of the block-updating procedure on
various quantities related to the scalar field theory.

\input{ref}

\end{document}

%% file: ref.tex
%